\documentclass[showpacs,preprintnumbers,twocolumn,
amsmath,amssymb,groupedaddress,superscriptaddress]{revtex4}
\usepackage{graphicx}
\usepackage{dcolumn}
\usepackage{bm}

\begin{document}

%\preprint{}

\title{Momentum distributions in medium and heavy exotic nuclei}

\author{M.~K.~Gaidarov}
\affiliation{Institute for Nuclear Research and Nuclear Energy,
Bulgarian Academy of Sciences, Sofia 1784, Bulgaria}

\author{G.~Z.~Krumova}
\affiliation{University of Ruse, Ruse 7017, Bulgaria}

\author{P.~Sarriguren}
\affiliation{Instituto de Estructura de la Materia, CSIC, Serrano
123, E-28006 Madrid, Spain}

\author{A.~N.~Antonov}
\affiliation{Institute for Nuclear Research and Nuclear Energy,
Bulgarian Academy of Sciences, Sofia 1784, Bulgaria}

\author{M.~V.~Ivanov}
\affiliation{Institute for Nuclear Research and Nuclear Energy,
Bulgarian Academy of Sciences, Sofia 1784, Bulgaria}

\author{E. Moya de Guerra}
\affiliation{Departamento de Fisica At\'omica, Molecular y
Nuclear, Facultad de Ciencias Fisicas, Universidad Complutense de
Madrid, E-28040 Madrid, Spain}

%\date{\today}

\begin{abstract}
We study nucleon momentum distributions of even-even isotopes of
Ni, Kr, and Sn in the framework of deformed self-consistent
mean-field Skyrme HF+BCS method, as well as of theoretical
correlation methods based on light-front dynamics and local
density approximation. The isotopic sensitivities of the
calculated neutron and proton momentum distributions are
investigated together with the effects of pairing and
nucleon-nucleon correlations. The role of deformation on the
momentum distributions in even-even Kr isotopes is discussed. For
comparison, the results for the momentum distribution in nuclear
matter are also presented.
\end{abstract}


\maketitle

\section{Introduction}

The study of nuclei close to the nuclear drip line and even beyond
it has been greatly extended in recent years. This increased
interest is based on new phenomena that already have been observed
or that are predicted to occur in these nuclei. Since the first
experiments \cite{Tan85a,Tan85b,Tan85c,Tan88a,Tan88b,Tan92}, it
has been found from analyses of total interaction cross sections
that weakly-bound neutron-rich nuclei, e.g. $^{6,8}$He, $^{11}$Li,
$^{14}$Be, $^{17,19}$B, have increased sizes that deviate
substantially from the $R\simeq A^{1/3}$ rule. It was realized
(e.g. Refs.~\cite{Han95,Dob94,Cas00}) that such a new phenomenon
is due to the weak binding of the last few nucleons which form a
diffuse nuclear cloud (the "nuclear halo") due to
quantum-mechanical penetration. Another effect is that the
nucleons can form a "neutron skin" \cite{Tan85c} when the neutrons
are on average less bound than the protons. The origin of the skin
lies in the large difference of the Fermi energy levels of protons
and neutrons so that the neutron wave function extends beyond the
effectively more bound proton wave functions \cite{Cas00}.

The experiments on scattering of radioactive nuclear beams on
proton target at various incident energies (e.g., less than 100
MeV/nucleon for He isotopes
\cite{Akop2001,Wolski2002,Giot2004,Rusek2001,Lap2001,Gil97,Lag2001,Gil96,Korsh97a,Korsh97b}
and 700 MeV/nucleon for He and Li isotopes
\cite{Alkhazov2002,Egel2001,Egel2002,Neum2002,Kiss2003}) have
allowed one to study the charge and matter distributions of these
nuclei using different phenomenological and theoretical methods
(e.g.,
\cite{Korsh97a,Korsh97b,Alkhazov2002,Egel2001,Egel2002,Neum2002,Kiss2003,Chulkov95,Crespo95,
Zhukov93,Zhukov94,Avrig2000,Avrig2002,Dort98,Kara97,Amos2006,Deb2005,Deb2001,Deb2003,Amos00,Arell2007}).

As known, the most accurate determination of the charge
distributions in nuclei can be obtained from electron-nucleus
scattering. For the case of the unstable exotic nuclei the
corresponding charge distributions are planned to be studied by
colliding electrons with these nuclei in storage rings (see, e.g.
the GSI physics program \cite{Shr01} and the plan of RIKEN
\cite{Sud01,Kat03}). A number of interesting issues can be
analyzed by the electron experiments. One of them is to study how
the charge distribution evolves with increasing neutron number at
fixed proton number, or to what extent the neutron halo or skin
may trigger sizable changes of the charge root-mean-squared (rms)
radius and the diffuseness in the peripheral region of the charge
distribution. In Ref.~\cite{Ant04} we studied charge form factors
of light exotic nuclei ($^{6,8}$He, $^{11}$Li, $^{14}$Be,
$^{17,19}$B) using various theoretical predictions of their charge
densities. Among the latter we used those from Tanihata {\it et
al.} \cite{Tan92} for the He isotopes, those from the
cluster-orbital shell-model approximation (COSMA) for the He
\cite{Zhukov94} and Li \cite{Korsh97a} isotopes, those from the
microscopic large-scale shell-model (LSSM) method (for He
\cite{Kar00} and Li \cite{Kar97}), and those from Suzuki {\it et
al.} \cite{Suz99} for $^{14}$Be and $^{17,19}$B nuclei. In
Ref.~\cite{Ant2005} our calculations of the charge form factors of
exotic nuclei were extended from light (He, Li) to medium and
heavy nuclei (Ni, Kr, and Sn). For the He and Li isotopes the
proton and neutron densities obtained in the LSSM method have been
used, while for Ni, Kr, and Sn isotopes the densities have been
obtained in the deformed self-consistent mean-field Skyrme
Hartree-Fock (HF)+BCS method \cite{Sar99,Guerra91,vautherin}. In
contrast to the work \cite{Ant04}, in \cite{Ant2005} we calculated
the charge form factors not only within the PWBA, but also in DWBA
by the numerical solution of the Dirac equation
\cite{Yen54,Luk02,Nis85} for electron scattering in the Coulomb
potential of the charge density of a given nucleus. The role of
the neutrons has been taken into account. A detailed study of the
charge radii and neutron skin in Ni, Kr, and Sn nuclei, as well as
of the formation of the proton skin, has been performed within the
same method in \cite{Sarriguren2007}.

Another important characteristic of the nuclear ground state is
the nucleon momentum distribution (NMD) $n(k)$. The scaling
analyses of inclusive electron scattering from a large variety of
nuclei (see, e.g. \cite{West80,CW99} for the $y$-scaling and
\cite{Alberico88,Barbaro98,Donnelly99a,Donnelly99b,Ant2004,Antonov2005,Ant2006a,Ant2006b}
for $\psi^{\prime}$-scaling and superscaling) showed the evidence
for the existence of high-momentum components of NMD at momenta
$k>2$ fm$^{-1}$. It has been shown
\cite{Ant2004,Antonov2005,Ant2006a,Ant2006b} that it is due to the
presence of nucleon-nucleon (NN) correlations in nuclei (for a
review, see e.g. \cite{Antbooks}). It has been pointed out that
this specific feature of $n(k)/A$ is similar for all nuclei, and
that it is a physical reason for the scaling and superscaling
phenomena in nuclei. As known \cite{Jaminon95,Antbooks}, the
mean-field approximation (MFA) is unable to describe
simultaneously the two important characteristics of the nuclear
ground state, the density and momentum distribution. Therefore, a
consistent analysis of the effects of the NN correlations on both
quantities is required using theoretical methods beyond the MFA in
the description of relevant phenomena, e.g. the scaling ones.
Particular attention has been paid to the NMD in a given
single-particle state analyzing the $(e,e^{\prime}p)$ reactions in
nuclei (see, e.g. the review in \cite{Frul84}, the works
\cite{Lapikas,Boffi82,Mougay76,Lanen93,Udias96} and references
therein). The self-consistent density-dependent HF (DDHF)
approximation has been applied in \cite{Guerra91} to calculate NMD
in spherical and deformed Nd isotopes, studying the effects of
deformation, as well as those of pairing and of dynamical
short-range NN correlations.

It is of importance to study the NMD not only in stable, but also
in exotic nuclei. It is known (see, e.g. Ref.~\cite{Thompson2001})
that in the reactions with exotic nuclei the momentum distribution
of a core fragment of the projectile reflects the momentum
distribution of the valence nucleons in the projectile near the
surface \cite{Tostevin98}. Many experimental (e.g, \cite{Korsh97})
and theoretical \cite{Bolen99} works have been carried out to
study the momentum distribution from the breakup of the
projectile. For instance, in Ref.~\cite{Horiuchi2007} the momentum
distribution of relative motion between two nucleons has been
calculated for both $^{6}$He and $^{6}$Li two-neutron halo nuclei.
The obtained results in the case of realistic NN interaction show
two interesting predictions: i) $S$-wave dominance in the NMD of
$^{6}$He and ii) the $^{6}$Li momentum distribution is very
similar to that of the deuteron. In \cite{Gaidarov96} the neutron
and proton momentum distributions in some stable nuclei ($^{12}$C,
$^{16}$O, $^{40}$Ca, $^{56}$Fe, and $^{208}$Pb) were calculated
along with those of light neutron-rich isotopes of Li, Be, B, and
C using the natural-orbital representation (NOR) on the basis of
the empirical data for $n(k)$ in $^{4}$He and, especially, those
for the high-momentum components of the latter.

The main aim of our work here is to calculate the NMD for the same
isotopic chains of neutron-rich nuclei (Ni, Kr, and Sn) for which
we had studied charge densities, radii, form factors, halo, and
skin in our previous works \cite{Ant2005} and
\cite{Sarriguren2007}. The mean-field contributions to $n(k)$ in
these nuclei are calculated within the same self-consistent
approach applied there in which the one-body energy density
functional is obtained starting from a two-body density-dependent
Skyrme interaction and a pairing interaction that is treated in
the BCS limit. The HF equations are solved  for the $(N,Z)$
nucleus using a deformed harmonic-oscillator basis in cylindrical
coordinates with oscillator lengths used as variational
parameters. The BCS equations  are solved at each HF iteration and
the occupation numbers are used to construct the density-dependent
mean field for the next HF iteration. We refer to this mean field
approach as DDHF+BCS. The remaining effects of the NN
interactions, to which we refer as NN correlations, are considered
in two of the correlations approaches, namely, in the approach
(see \cite{AGI+02,Antonov2005,Ant2006b}) using the light-front
dynamics (LFD) method (e.g., \cite{Carbonell95}) and in that
\cite{Strin90} based on the local density approximation (LDA).
Several questions are investigated, such as the sensitivity of
$n(k)$ to all details of the calculations, e.g.: i) to different
types of Skyrme forces; ii) to the pairing correlation effects;
iii) to the effects of nuclear deformation; iv) to the strength of
the NN correlations included in the LFD and LDA approaches
(respectively, to the values of the correlations strength
parameters $\beta$ and $\gamma$). A special attention is paid to
the isotopic and isotonic sensitivity of the proton and neutron
momentum distributions. The results for $n(k)$ in the exotic
nuclei are compared with that in nuclear matter (NM).

The paper is organized in the following way. Section
\ref{section2} contains the formalism of the deformed Skyrme
HF+BCS method and the approaches based on the LFD and LDA methods
that provide the model nucleon momentum distributions. The
numerical results and discussions are presented in
Sec.~\ref{section3}. Finally, we draw the main conclusions of this
study in Sec.~\ref{summary}.

\section{Theoretical framework}\label{section2}

\subsection{Deformed Skyrme HF+BCS formalism}

Some of the results discussed in the next Section have been
obtained from self-consistent deformed Hartree-Fock calculations
with density dependent Skyrme interaction \cite{vautherin} and
pairing correlations. Pairing between like nucleons has been
included by solving the BCS equations at each iteration either
with a fixed pairing gap parameter (determined from the odd-even
experimental mass differences) or with a fixed pairing strength
parameter. We consider in this paper the Skyrme force SLy4
\cite{sly4} which gives an appropriate description of bulk
properties of spherical and deformed nuclei.

Assuming time reversal invariance, the single-particle
Hartree-Fock solutions for axially symmetric deformed nuclei are
characterized by the eigenvalue $\Omega_i$ of the third component
of the total angular momentum on the symmetry axis and by the
parity $\pi_i$. The state $i$ can be written as
\begin{eqnarray}
\Phi_i \left( {\vec r},\sigma ,q\right)& = &\chi_{q_i}(q) \big[
\Phi^+_i (r_{\perp},z) e^{i\Lambda^-\varphi} \chi_+(\sigma) \nonumber \\
&+& \Phi^-_i (r_{\perp},z) e^{i\Lambda^+\varphi} \chi_-(\sigma) \big ]\, ,
\end{eqnarray}
where $\chi_{q_i}(q)$, $\chi_{\pm}(\sigma)$ are isospin and spin
functions, $\Lambda^{\pm}=\Omega_i \pm 1/2 \ge 0$. $r_{\perp},z,\varphi$
are the cylindrical coordinates of ${\vec r}$.

The wave functions $\Phi_i$ are expanded into the eigenfunctions
$\phi_{\alpha}$ of an axially symmetric deformed
harmonic-oscillator potential in cylindrical coordinates. We use
12 major shells in this expansion,
\begin{equation}
\Phi_i \left( {\vec r},\sigma ,q\right) = \chi_{q_i}(q)
\sum_{\alpha} C^i_{\alpha} \phi_{\alpha}\left( {\vec r},\sigma
\right) \,  ,
\label{eq:expandmom}
\end{equation}
with $\alpha=\{n_{\perp},n_z,\Lambda,\Sigma\}$ and
\begin{equation}
\phi_{\alpha}\left( {\vec r},\sigma \right)=
\psi^{\Lambda}_{n_{\perp}}(r_{\perp}) \psi_{n_z}(z) \frac{e^{i\Lambda
\varphi}}{\sqrt{2\pi}} \chi_{_\Sigma}(\sigma)\, ,
\end{equation}
in terms of Hermite and Laguerre polynomials
\begin{equation}
\psi_{n_z}(z)= \sqrt{\frac{1}{\sqrt{\pi}2^{n_z}n_z!}} \,
\beta^{1/2}_z\, e^{-{\xi}^2/2}\, H_{n_z}(\xi) \, ,
\end{equation}
\begin{equation}
\psi^{\Lambda}_{n_{\perp}}(r_{\perp})=\sqrt{\frac{n_{\perp}}
{(n_{\perp}+\Lambda )!}} \,
\beta_{\perp}\, \sqrt{2}\, \eta^{\Lambda/2}\, e^{-\eta/2}\,
L_{n_{\perp}}^{\Lambda}(\eta) \,
\end{equation}
with
\begin{eqnarray}
\beta_z=(m\omega_z/\hbar )^{1/2}&,&\quad
\beta_\perp=(m\omega_\perp/\hbar )^{1/2},\nonumber \\
\quad \xi=z\beta_z&,&\quad \eta=r_{\perp}^2\beta_\perp ^2 \, .
\end{eqnarray}

From the expansion (\ref{eq:expandmom}) we may conveniently
express the single-particle Hartree-Fock wave functions in
momentum space, which we denote as $\tilde{\Phi}_i ( {\vec
k},\sigma ,q)$. They are given by \cite{Guerra91}
\begin{equation}
\tilde{\Phi}_i ({\vec k},\sigma ,q) = \chi_{q_i}(q) \sum_{\alpha}
C^i_{\alpha} \tilde{\phi}_{\alpha}({\vec k},\sigma ) \,
\end{equation}
with
\begin{equation}
\tilde{\phi}_{\alpha}({\vec k},\sigma )=\frac{1}{(2\pi)^{3/2}}
\int d{\vec r}e^{-i{\vec k}{\vec r}}\phi_{\alpha}({\vec r},\sigma)
\label{eq:wfmom}
\end{equation}
normalized to one.

The spin-independent proton, neutron and total densities are given
by
\begin{equation}
\rho({\vec r})=\rho (r_{\perp},z)=\sum _{i} 2v_i^2\rho_i(r_{\perp},z)\, ,
\end{equation}
in terms of the occupation probabilities $v_i^2$ resulting from
the BCS equations and the single-particle densities $\rho_i$
\begin{equation}
\rho_i({\vec r})=  \rho_i(r_{\perp},z)=|\Phi^+_i(r_{\perp},z)|^2+
|\Phi^-_i(r_{\perp},z)|^2 \, ,
\end{equation}
with
\begin{eqnarray}
\Phi^\pm _i(r_{\perp},z)&=&{1\over \sqrt{2\pi}}\nonumber \\
&\times & \sum_{\alpha}\, \delta_{\Sigma, \pm 1/2}\,
\delta_{\Lambda,\Lambda^\mp}\, C_\alpha ^i\, \psi_{n_{\perp}}^\Lambda
(r_{\perp}) \, \psi_{n_z}(z) \, .
\end{eqnarray}
Similarly, we define in momentum space the proton, neutron and
total momentum distributions by
\begin{equation}
n({\vec k})=n(k_{\perp},k_{z})=\sum _{i} 2v_i^2n_i(k_{\perp},k_{z})\, ,
\end{equation}
where $k_{\perp},k_z$ are the cylindrical coordinates of ${\vec k}$.
The single-particle momentum distributions $n_{i}({\vec k})$ are given by
\begin{equation}
n_{i}({\vec k})=n_i(k_{\perp},k_{z})=|\tilde{\Phi}^+_i(k_{\perp},k_{z})|^2+
|\tilde{\Phi}^-_i(k_{\perp},k_{z})|^2\, ,
\end{equation}
where
\begin{eqnarray}
\tilde{\Phi}^\pm _i(k_{\perp},k_{z})&=&{1\over \sqrt{2\pi}} \\
&\times & \sum_{\alpha} \delta_{\Sigma, \pm 1/2}
\delta_{\Lambda,\Lambda^\mp} C_\alpha ^i (-i)^{N}
\psi_{n_{\perp}}^\Lambda (k_{\eta}) \psi_{n_z}(k_{\xi}), \nonumber
\end{eqnarray}
\begin{equation}
k_{\xi}=k_{z}/\beta_{z}, \,\,\,\,\,
k_{\eta}=k_{\perp}^{2}/\beta_{\perp}^{2}=\frac{k_{x}^{2}+k_{y}^{2}}{\beta_{\perp}^{2}},
\end{equation}
and $N=\Lambda+n_{z}+2n_{\perp}$ is the major shell quantum number of
the basis state $\alpha$.

The multipole decomposition of the momentum distribution can be
written as
\begin{eqnarray}
n({\vec k}) & = &\sum_{\lambda} n_{\lambda}(k)
P_{\lambda}(\cos\theta_{k})\nonumber \\
&=& n_0(k) + n_2(k)\, P_2(\cos \theta_{k}) + \ldots \,
\label{nmult}
\end{eqnarray}
with multipole components $\lambda$
\begin{eqnarray}
n_{\lambda}(k)&=&\frac{2\lambda +1}{2}
\int_{-1}^{+1} P_{\lambda}(\cos\theta_{k})\nonumber \\
&\times & n(k\cos\theta_{k},k\sin\theta_{k}) d(\cos\theta_{k}).
\end{eqnarray}

\subsection{Methods going beyond the mean-field approximation}

It is well known (e.g. \cite{Antbooks}) that the methods within
the MFA (e.g. shell model, Hartree-Fock and others) can describe
the nucleon momentum distribution $n(k)$ only for momentum values
$k<$1.5 fm$^{-1}$ and are unable to explain $n(k)$ for larger $k$.
The high-momentum components of $n(k)$ ($k>$1.5 fm$^{-1}$) are due
to the specific forces between the nucleons near the nuclear core
($r_{c}\approx 0.4$ fm) that are the reasons for the short-range
and tensor NN correlations. The differences between the values of
$n(k)$ for large $k$ obtained within the correlation methods
(e.g., $\exp$($S$)-method \cite{Zab78}, the Jastrow correlation
method and others, for a review see e.g. \cite{Antbooks}) reach
orders of magnitude. In this Subsection we consider the effects of
NN correlations included in two correlation methods on the
high-momentum contributions to the nucleon momentum distribution.

\subsubsection{Theoretical approach based on the light-front dynamics method}

Here we derive the NMD obtained within the LFD approach developed
in Refs.~\cite{AGI+02,Antonov2005}. The latter is based on the
nucleon momentum distribution in the deuteron from the light-front
dynamics method (e.g. \cite{Carbonell95}). Using the
natural-orbital representation of the one-body density
matrix~\cite{Low55}, $n(k)$ is written as a sum of the
contributions from the hole-states $[{\tilde n}^{h}(k)]$ (states
up to the Fermi level (F.L.)) and the particle-states $[{\tilde
n}^{p}(k)]$ (see also~\cite{Antonov2005}) for protons (Z) and
neutrons (N):
\begin{equation}
n_{Z(N)}(k)={\tilde n}_{Z(N)}^{h}(k)+ {\tilde n}_{Z(N)}^{p}(k),
\label{eq:35}
\end{equation}
where
\begin{equation}
{\tilde n}_{Z(N)}^{h}(k)=\frac{C(k)}{Z(N)}\sum_{nlj}^{F.L.}
2(2j+1)\lambda_{nlj}|{\tilde R}_{nlj}(k)|^{2}
\label{eq:35a}
\end{equation}
and
\begin{equation}
{\tilde n}_{Z(N)}^{p}(k)=\frac{C(k)}{Z(N)}\sum_{F.L.}^{\infty}
2(2j+1)\lambda_{nlj}|{\tilde R}_{nlj}(k)|^{2}.
\label{eq:35b}
\end{equation}
In (\ref{eq:35a}) and (\ref{eq:35b})
\begin{equation}
C(k)= \frac{m_N}{(2\pi)^3\sqrt{k^2+m_N^2}},
\label{eq:36}
\end{equation}
$m_N$ being the nucleon mass, $\lambda_{nlj}$ are the natural
occupation numbers and ${\tilde R}_{nlj}(k)$ are the corresponding
wave functions in $k$-space  of protons (neutrons) in states with
quantum numbers $nlj$ . The momentum distribution (\ref{eq:35}) is
normalized to unity:
\begin{equation}
\int n_{Z(N)}(k)d{\vec k}=1.
\label{eq:35c}
\end{equation}

As shown in \cite{AGI+02}, in the NOR the hole-state contribution
${\tilde n}_{Z(N)}^{h}(k)$ to the momentum distribution can be
represented to a good approximation  by the momentum distribution
from the spherical mean-field methods, i.e. $\lambda_{nlj}$ are
taken to be unity and ${\tilde R}_{nlj}(k)$ are replaced by the
corresponding mean-field eigenfunctions. In our work we substitute
${\tilde n}_{Z(N)}^{h}(k)$ in (\ref{eq:35}) and (\ref{eq:35a}) by
\begin{equation}
{\tilde n}_{Z(N)}^{h}(k)=\frac{C(k)}{Z(N)}{\tilde n}_{Z(N)}(k),
\label{eq:37}
\end{equation}
where ${\tilde n}_{Z(N)}(k)$ is expressed by the NMD
$n^{DDHF}_{Z(N)}(k)$ obtained within the DDHF formalism (see
Sec.~II.A.)
\begin{equation}
{\tilde n}_{Z(N)}(k)=\frac{Z(N)n^{DDHF}_{Z(N)}(k)} {\int d{\vec
k}^{\prime}C(k^{\prime})n^{DDHF}_{Z(N)}(k^{\prime})}
\label{eq:37b}
\end{equation}
and the normalization is
\begin{equation}
\int C(k) {\tilde n}_{Z(N)}(k) d{\vec k}=Z(N).
\label{eq:37a}
\end{equation}
We note that ${\tilde n}^{h}_{Z(N)}(k)$, ${\tilde
n}^{p}_{Z(N)}(k)$ and ${\tilde n}_{Z(N)}(k)$ include the function
$C(k)$ that originates from the relativistic LFD approach.

Concerning the particle-state contribution [${\tilde
n}_{Z(N)}^{p}(k)$] in (\ref{eq:35}) and (\ref{eq:35b}) we used in
Refs.~\cite{AGI+02,Antonov2005} the well-known facts that: (i) the
high-momentum components of $n(k)$ caused by short-range and
tensor correlations are almost completely determined by the
contributions of the particle-state natural orbitals
(e.g.~\cite{SAD93}), and (ii) the high-momentum tails of the
momentum distributions per particle are approximately equal for
all nuclei and are a rescaled version of the nucleon momentum
distribution in the deuteron $n_d(k)$~\cite{FCW00,Ciofi96},
\begin{equation}
n_A(k)\simeq C^{A} n_d(k),
\label{eq:38}
\end{equation}
where $C^{A}$ is a constant. These facts made it possible to
assume in~\cite{AGI+02,Antonov2005} and later, using the
modification of the approach in \cite{Ant2006b}, that ${\tilde
n}_{Z(N)}^{p}(k)$ is related to the high-momentum components of
the nucleon momentum distributions in the deuteron. Thus ${\tilde
n}_{Z(N)}^{p}(k)$ from Eq.~(\ref{eq:35b}) can be substituted by
(up to a normalization factor):
\begin{equation}
{\tilde n}_{Z(N)}^{\text{p}}(k)=\beta \big[n_2(k)+n_5(k)\big],
\label{eq:39}
\end{equation}
where $\beta$ is a parameter, and $n_2(k)$ and $n_5(k)$ are
expressed by angle-averaged functions~\cite{AGI+02} as:
\begin{equation}
n_2(k)= C(k) \overline{f_2^2(k)}
\label{eq:40a}
\end{equation}
and
\begin{equation}
n_5(k)= C(k) \overline{(1-z^2) f_5^2(k)}.
\label{eq:40b}
\end{equation}
In~(\ref{eq:40b}) $z=\cos(\widehat{\vec{k},\vec{n}})$, $\vec{n}$
being a unit vector along the three-vector ($\vec{\omega}$)
component of the four-vector $\omega$ which determines the
position of the light-front surface~\cite{Carbonell95}. The
functions $f_2(k)$ and $f_5(k)$ are two of the six scalar
functions $f_{1-6}(k^2,\vec{n}\cdot\vec{k})$ which are the
components of the deuteron total wave function
$\Psi(\vec{k},\vec{n})$. It was shown~\cite{Carbonell95} that
$f_5$ largely exceeds other $f$-components for $k\geq$ 2.0--2.5
fm$^{-1}$ and is the main contribution to the high-momentum
component of $n_d(k)$, incorporating the main part of the
short-range properties of the NN interaction. It was shown in
\cite{Ant2006b}, however, that not only $n_5(k)$ (originating from
$f_5(k)$ (\ref{eq:40b})), but also $n_2(k)$ (related to $f_2(k)$
(\ref{eq:40a})) has to be included partially in the particle-state
contribution ${\tilde n}_{Z(N)}^{p}(k)$ to the momentum
distribution. The latter leads to a successful explanation of the
experimental data for the quasielastic scaling function
$f^{QE}(\psi^{\prime})$ (see Fig.~4 of Ref.~\cite{Ant2006b})
extracted from inclusive electron scattering off nuclei.

Finally, the normalized to unity proton (neutron) momentum
distribution [Eqs.~(\ref{eq:35})--(\ref{eq:35c})] has the form:
\begin{widetext}
\begin{equation}
n_{Z(N)}(k)=\dfrac{C(k) {\tilde n}_{Z(N)}(k)+Z(N)
\beta\big[n_2(k)+n_5(k)\big]}{\int d{\vec
k^{\prime}}\Big\{C(k^{\prime}) {\tilde
n}_{Z(N)}(k^{\prime})+Z(N)\beta\big[n_2(k^{\prime})+n_5(k^{\prime})\big]\Big\}}
\label{eq:41}
\end{equation}
\end{widetext}
(with normalization of ${\tilde n}_{Z(N)}(k)$ presented by
Eq.~(\ref{eq:37a})).

For the value of the parameter $\beta$ we choose $\beta=5.0$
because of three reasons. Firstly, our calculations of
$n_{Z(N)}(k)$ for various nuclei using this value lead to results
very close to the empirical data for the nucleon momentum
distribution $n_{CW}$ (see Fig.~2 of Ref.~\cite{Antonov2005} and
Fig.~3 of Ref.~\cite{Ant2006b}) extracted in
Refs.~\cite{CW99,CW97} from the $y$-scaling analyses of inclusive
electron scattering off nuclei. Secondly, our estimations of the
high-momentum components of $n_{Z(N)}(k)$ [Eq.~(\ref{eq:41})]
showed that the value of $\beta$ must be similar to the value of
$C^{A}$ [Eq.~(\ref{eq:38})]. It is shown (see Table~I in
\cite{Ciofi96}) that the value of $C^{A}$ estimated from the
height of the plateau exhibited by the ratio of the nucleon
momentum distribution of a nucleus to the one of the deuteron at
$k>$ 2.0 fm$^{-1}$ ranges from 4.0 for $^{12}$C to 4.4 for
$^{40}$Ca, 4.5 for $^{56}$Fe, 4.8 for $^{208}$Pb, and 4.9 for
nuclear matter ($A$=$\infty$). Third, our results for the NMD
(e.g. for $^{12}$C, $^{64}$Ni and others) in LFD approach for
large values of $k$ ($k>$ 2.0 fm$^{-1}$) are similar to those
obtained in the local density approximation using the Jastrow
correlation method \cite{Strin90}.

\subsubsection{Theoretical approach based on the local density
approximation}

It is well known that the inclusion of correlations in nuclear
matter modifies the occupation probability predicted by the local
density Fermi gas model. It was shown in \cite{Strin90} that for a
finite nucleus the separation between the mean-field contribution
and correlation effects can be performed in an analogous way.
According to Ref.~\cite{Strin90} one can introduce proton
(neutron) momentum distribution in a general way:
\begin{equation}
n_{Z(N)}(k)=n^{MFA}_{Z(N)}(k)+\delta n_{Z(N)}(k),
\label{eq:ldamomdis}
\end{equation}
where $n^{MFA}_{Z(N)}(k)$ is the mean-field contribution which can
be momentum distribution corresponding to a Slater determinant
generated by the single-particle wave functions in the momentum
space or HF momentum distribution, while $\delta n_{Z(N)}(k)$
embodies the corrections due to dynamical correlations not
included in the MFA. If one applies the LDA to the second term of
Eq.~(\ref{eq:ldamomdis}), the nucleon momentum distribution
$n_{Z(N)}(k)$ can be written in the form:
\begin{equation}
n_{Z(N)}(k)=n^{MFA}_{Z(N)}(k)+\frac{1}{4\pi^{3}} \int \delta \nu
(k_{F}^{Z(N)}(r),k)d{\vec r},
\label{eq:momdis}
\end{equation}
where $\delta \nu (k_{F}^{Z(N)}(r),k)$ corresponds to the
occupation probability that is entirely due to the effects of
dynamical correlations induced by the NN interaction. The local
Fermi momentum $k_{F}^{Z(N)}(r)$ is related to the proton
(neutron) density through the relation
\begin{equation}
k_{F}^{Z(N)}(r)=\left [3\pi^{2}\rho_{Z(N)}(r)\right ]^{1/3}.
\label{eq:kf}
\end{equation}
By definition of $k_{F}^{Z(N)}(r)$ one has $\int \delta \nu
(k_{F}^{Z(N)}(r),k) d{\vec k}=0$. For convenience, a
phenomenological procedure based on the results of the lowest
order cluster (LOC) approximation developed in \cite{Flynn84} has
been followed to evaluate explicitly the correlated term. Choosing
a correlation function of the form
\begin{equation}
f(r)=1-e^{-\gamma^{2}r^{2}},
\label{eq:corrfunc}
\end{equation}
the LOC gives for $\delta \nu(k_{F}^{Z(N)},k)$ \cite{Strin90}
\begin{eqnarray}
\delta \nu (k_{F}^{Z(N)}(r),k)&=&\left [Y(k,8)-k_{dir}\right
]\Theta(k_{F}^{Z(N)}(r)-k)\nonumber \\
&+&8\left \{k_{dir}Y(k,2)-[Y(k,4)]^{2}\right \},
\label{eq:corrterm}
\end{eqnarray}
where
\begin{eqnarray}
c_{\mu}^{-1}Y(k,\mu)&=&
\frac{e^{-\tilde{k}_{+}^{2}}-e^{-\tilde{k}_{-}^{2}}}{2\tilde{k}}+
\int_{0}^{\tilde{k}_{+}}e^{-y^{2}}dy\\
&+& {\text{sgn}}
(\tilde{k}_{-})\int_{0}^{|\tilde{k}_{-}|}e^{-y^{2}}dy \nonumber
\label{eq:ygrek}
\end{eqnarray}
with
\begin{equation}
c_{\mu}=\frac{1}{8\sqrt{\pi}}\left(\frac{\mu}{2}\right
)^{3/2},\;\;\; \tilde{k}=\frac{k}{\gamma\sqrt{\mu}},
\label{eq:constants1}
\end{equation}
\begin{equation}
\tilde{k}_{\pm}=\frac{k_{F}\pm k}{\gamma\sqrt{\mu}},\;\;\;
{\text{sgn}}(x)=\frac{x}{|x|}. \label{eq:constants2}
\end{equation}
The quantity
\begin{eqnarray}
k_{dir}&=&\frac{2\big[k_{F}^{Z(N)}\big]^{3}}{3\pi^{2}}\int\big[f(r)-1\big]^{2}d{\vec
r}\nonumber \\
&=&\frac{1}{3\sqrt{2\pi}}\left(\frac{k_{F}^{Z(N)}}{\gamma}\right)^{3}
\label{eq:jaspar}
\end{eqnarray}
is the direct part of the Jastrow wound parameter.

As in the approach based on the LFD method discussed before, in
our work we use for $n^{MFA}_{Z(N)}(k)$ the momentum distributions
obtained from DDHF calculations $n_{Z(N)}^{DDHF}(k)$ with
normalization given by:
\begin{equation}
\int n_{Z(N)}^{DDHF}(k) d{\vec k}=Z(N).
\label{eq:normn}
\end{equation}
For the densities $\rho_{Z(N)}(r)$ entering Eq.~(\ref{eq:kf}) we
use the HF+BCS proton (neutron) densities \cite{Sarriguren2007}
normalized as:
\begin{equation}
\int \rho_{Z(N)}^{DDHF}(r) d{\vec r}=Z(N).
\label{eq:normrho}
\end{equation}
For the correlation factor in Eq.~(\ref{eq:corrfunc}) we adopt the
same value $\gamma$=1.1 fm$^{-1}$ as in Ref.~\cite{Strin90} which
is taken from the microscopic nuclear matter calculations
\cite{Fant84} but seems to agree also with the data on momentum
distributions $n(k)$ in finite nuclei.

\section{Results of calculations and discussion}
\label{section3}

\begin{figure*}
\centering
\includegraphics[width=150mm]{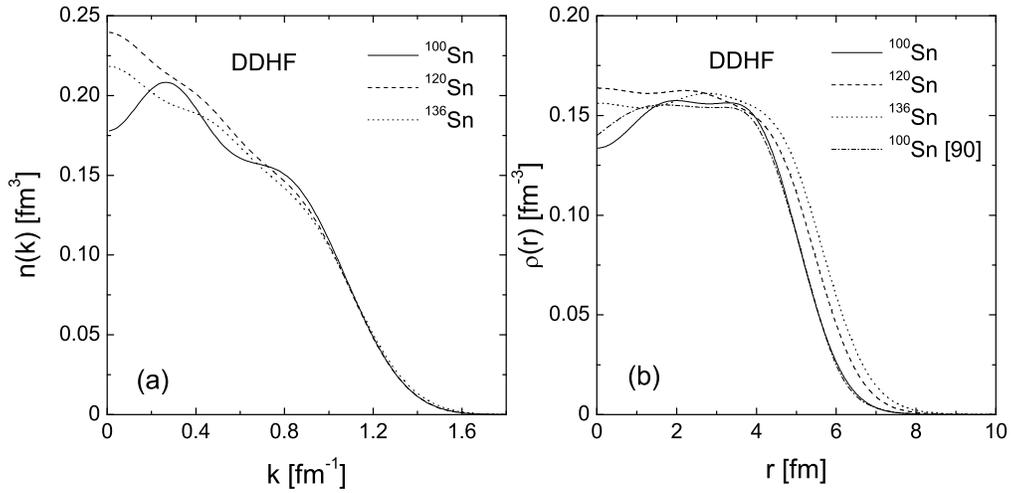}
\caption[]{(a) Total momentum distribution for $^{100}$Sn,
$^{120}$Sn, and $^{136}$Sn isotopes resulting from DDHF
calculations. The normalization is: $\int  n(k) d{\vec k}=1$; (b)
Total density distributions for the same Sn isotopes. The
normalization is: $\int  \rho(r) d{\vec r}=A$.
\label{fig1}}
\end{figure*}

We start by showing the results of our calculations for the
monopole components of $n(\vec k)$ and $\rho(\vec r)$ in the
$^{100,120,136}$Sn isotopes. As discussed in
Ref.~\cite{Guerra91,Sarriguren2007} these are the only important
components for the HF momentum distributions ($n(\vec k)\cong
n_{0}(\vec k)\equiv n(k)$ [Eq.~(\ref{nmult})]). For easier
reading, in the figures we will omit the subscript $Z(N)$-indices
of $n(k)$ for the proton and neutron momentum distributions
replacing them by $p(n)$, respectively. In Fig.~\ref{fig1}a are
given the total momentum distributions for $^{100}$Sn, $^{120}$Sn
and $^{136}$Sn plotted in non-logarithmic scale. One can observe
in the figure an appreciable difference between the curves in the
low momentum region and, at the same time on this scale, no
sensitivity of $n(k)$ at $k>1$ fm$^{-1}$ when increasing the
number of neutrons. As can be expected, the momentum distributions
calculated within the DDHF approach show a steep decrease up to
$k\leq 2$ fm$^{-1}$ that can be seen in all mean-field
calculations. Here we emphasize that the effects of NN
correlations on the momentum distributions can be seen at higher
momenta ($k\geq 2$ fm$^{-1}$) when we will compare the DDHF
results with those obtained in the approaches which take them into
account. The results for $n(k)$ are related with the total density
profiles of the selected three Sn isotopes shown in
Fig.~\ref{fig1}b. For a comparison, in the same figure we present
the total density $\rho(r)$ of $^{100}$Sn obtained in
Ref.~\cite{Amos2004} within the Hartree-Fock-Bogoliubov model
using the SLy4 parameterization of the Skyrme force. We note,
following the discussion concerning Fig.~5 in
\cite{Sarriguren2007}, that the effect of adding more and more
neutrons leads to an extension of the total density $\rho(r)$ as
one goes from $^{100}$Sn to $^{136}$Sn and to an emergence of a
region at the nuclear surface quantified as a "neutron skin". The
latter is due to the larger spatial extension of the neutron
density relatively to the proton one. The same consistent trends
of the matter densities when increasing the mass number $A$ were
found in \cite{Amos2004}. One can see from Fig.~\ref{fig1}b that
the densities of $^{100}$Sn obtained in our previous work
\cite{Sarriguren2007} and in Ref.~\cite{Amos2004} are very
similar. Second, the same choice of two extreme neutron-deficient
and neutron-rich isotopes and one stable isotope between them in
the Sn chain \cite{Sarriguren2007} makes the difference in the
results for $n(k)$ at low $k$ more pronounced with respect to
those for $\rho(r)$ at large $r$.

\begin{figure*}
\centering
\includegraphics[width=150mm]{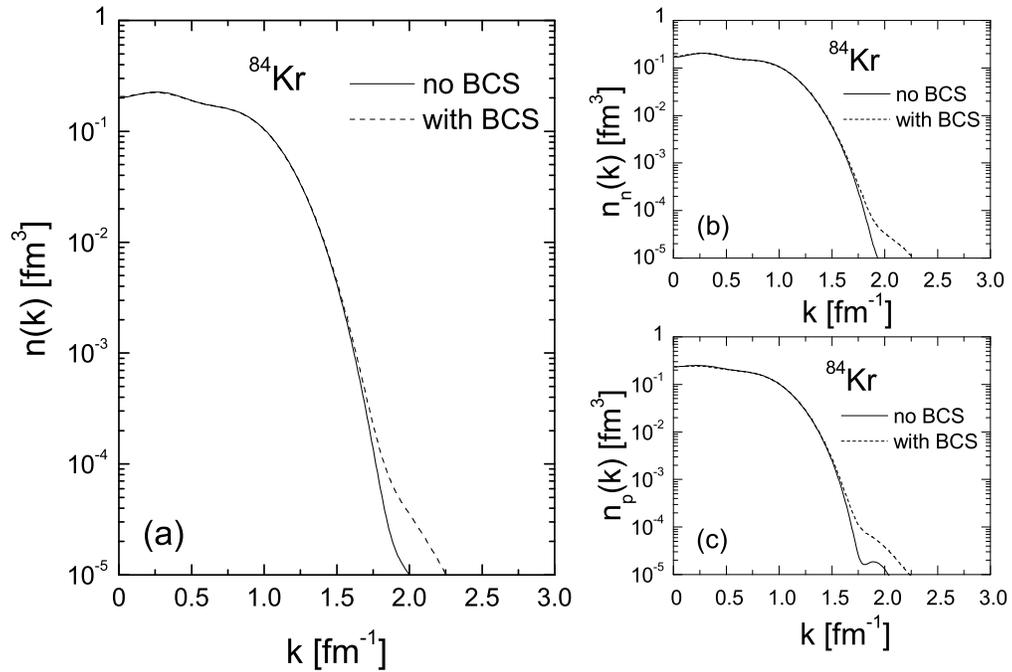}
\caption[]{DDHF results for the total (a), neutron (b), and proton
(c) momentum distributions with (dashed line) and without (solid
line) pairing for $^{84}$Kr. The normalization is: $\int
n_{n(p)}(k)d{\vec k}=\int n(k)d{\vec k}=1$. \label{fig2}}
\end{figure*}

The effect of pairing correlations on the momentum distribution
can be seen in Fig.~\ref{fig2}. We restrict the discussion to the
stable $^{84}$Kr isotope which is weakly deformed. In this case
the pairing correlations have been included by solving the BCS
equations in the constant pairing gap mode. Fig.~\ref{fig2} shows
that the effect of pairing correlations is small. At high $k$ this
effect plays some role resulting in more pronounced tails when
BCS-correlations are included in the calculations. Nevertheless,
as we will illustrate and discuss later, the effects of pairing
correlations on the HF momentum distribution of nuclear matter and
of finite nuclei are very different being much larger in the case
of NM. Since there is no big difference between the results for
$n_{p}(k)$, $n_{n}(k)$, and $n(k)$ with or without
BCS-correlations included, afterwards we will use in our
consideration only the results from the complete DDHF+BCS
calculations.

\begin{figure*}
\centering
\includegraphics[width=160mm]{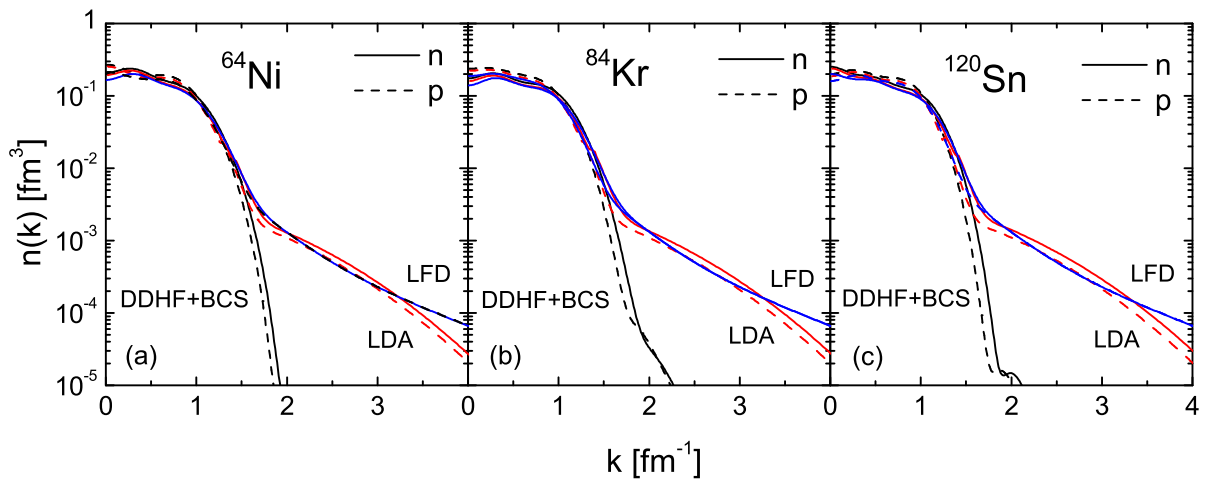}
\caption[]{Neutron (solid line) and proton (dashed line) momentum
distributions obtained within the DDHF+BCS (black), LFD (blue),
and LDA (red) methods for $^{64}$Ni (a), $^{84}$Kr (b), and
$^{120}$Sn (c) nuclei. The normalization is: $\int
n_{n(p)}(k)d{\vec k}=1$. \label{fig3}}
\end{figure*}

Our next step is to present and discuss the results for the NMD's
of exotic nuclei obtained also in correlation methods. A
comparison of these results for the neutron and proton momentum
distributions of $^{64}$Ni, $^{84}$Kr, and $^{120}$Sn nuclei is
given in Fig.~\ref{fig3} together with the HF momentum
distributions. As can be seen, for all nuclei the inclusion of NN
correlations strongly affects the high-momentum region of NMD. At
$k>1.5$ fm$^{-1}$ both LFD and LDA momentum distributions start to
deviate from the DDHF+BCS case. They behave rather similar in the
interval $1.5<k<3$ fm$^{-1}$. At $k>3$ fm$^{-1}$ the LFD method
predicts systematically higher momentum components compared to LDA
momentum distributions. This observation can be explained by the
different extent to which NN correlations are taken into account
in both approaches. Our results for the NMD's in the LFD method
for large values of $k$ ($k>2$ fm$^{-1}$) are similar to those
obtained within the Jastrow correlation method and, thus, the
high-momentum tails of $n(k)$ are caused by the short-range NN
correlations. The LDA approach through the nuclear matter dynamic
effects and using the local Fermi momentum $k_{F}^{Z(N)}(r)$
calculated self-consistently by means of the HF density
[Eq.~(\ref{eq:kf})] produces less pronounced high-momentum tail,
but still the results are very close to those obtained in the LFD
method. As was already shown, at $k>1.5$ fm$^{-1}$ the DDHF+HF
momentum distributions fall off rapidly by several orders of
magnitude in contrast to the correlated NMD's. In addition, we
observe that: i) the results shown above are similar for all
nuclei in a given isotopic chain and going from Ni to Sn isotopes,
as well; ii) the behavior of $n(k)$ is similar for protons and
neutrons; iii) at high $k$ the proton and neutron NMD's obtained
within the LFD method cannot be distinguished from each other
because the high-momentum tails in this approach are determined by
the high-momentum component of the nucleons in the deuteron
\cite{AGI+02}; iv) concerning the NMD's calculated in the LDA
approach, some difference between $n(k)$ for protons and neutrons
can be observed due to $Z(N)$-dependence of the local Fermi
momentum $k_{F}$.

In the next figures \ref{fig4}-\ref{fig7} we show the neutron
$n_{n}(k)$ [Figs.~\ref{fig4} and \ref{fig6}] and proton $n_{p}(k)$
[Figs.~\ref{fig5} and \ref{fig7}] momentum distributions of some
selected isotopes in the Ni and Sn chains, the same which have
been considered in Ref.~\cite{Sarriguren2007} to calculate
important nuclear properties in coordinate space. The results are
presented in both logarithmic and linear scales in order to study
the isotopic sensitivity of these momentum distributions in the
high-momentum region and in the region of small momenta,
respectively. In addition, in each of the figures the results for
neutron and proton momentum distributions in the DDHF+BCS method
and in the correlation LFD and LDA approaches at $k<2$ fm$^{-1}$
are given separately in panels (b) and (c). First, it is seen from
the figures that the general trend in the behavior of NMD's
obtained within the methods used in the calculations and already
shown in Fig.~\ref{fig3} is preserved. Second, the evolution of
the NMD's as we increase the number of neutrons consists of an
increase of the high-momentum tails (for $k>1.5$ fm$^{-1}$) of
$n_{n}(k)$, while the effect on $n_{p}(k)$ is opposite. However,
the spreading of the tails corresponding to $n_{p}(k)$ of the
considered isotopes is of the same order although the number of
protons remains the same. In this respect, the results presented
in Figs.~\ref{fig5} and \ref{fig7} are challenging because they
show how proton momentum distributions "feel" the different number
of neutrons in exotic nuclei. We would like also to emphasize that
the LFD method does not show this isotopic sensitivity, in
contrast to the HF and LDA methods which still demonstrate this
trend. Concerning the low-momentum region it can be seen from
Figs.~\ref{fig4}-\ref{fig7} that NMD's are very sensitive to the
details of the calculations. In this region $n_{n}(k)$ decreases
while, on the contrary, $n_{p}(k)$ increases with the increase of
the number of neutrons $N$. This is a common feature of the
calculated results obtained in all methods. Nevertheless, in this
region the spreading is considerably reduced. From the comparison
of the proton and neutron momentum distributions at low momenta it
comes out that the protons exhibit the same trend
[Figs.~\ref{fig5} and \ref{fig7}((b) and (c))], while the neutrons
are more sensitive to nuclear shell effects (see panels (b) and
(c) of Figs.~\ref{fig4} and \ref{fig6}).

\begin{figure*}
\centering
\includegraphics[width=150mm]{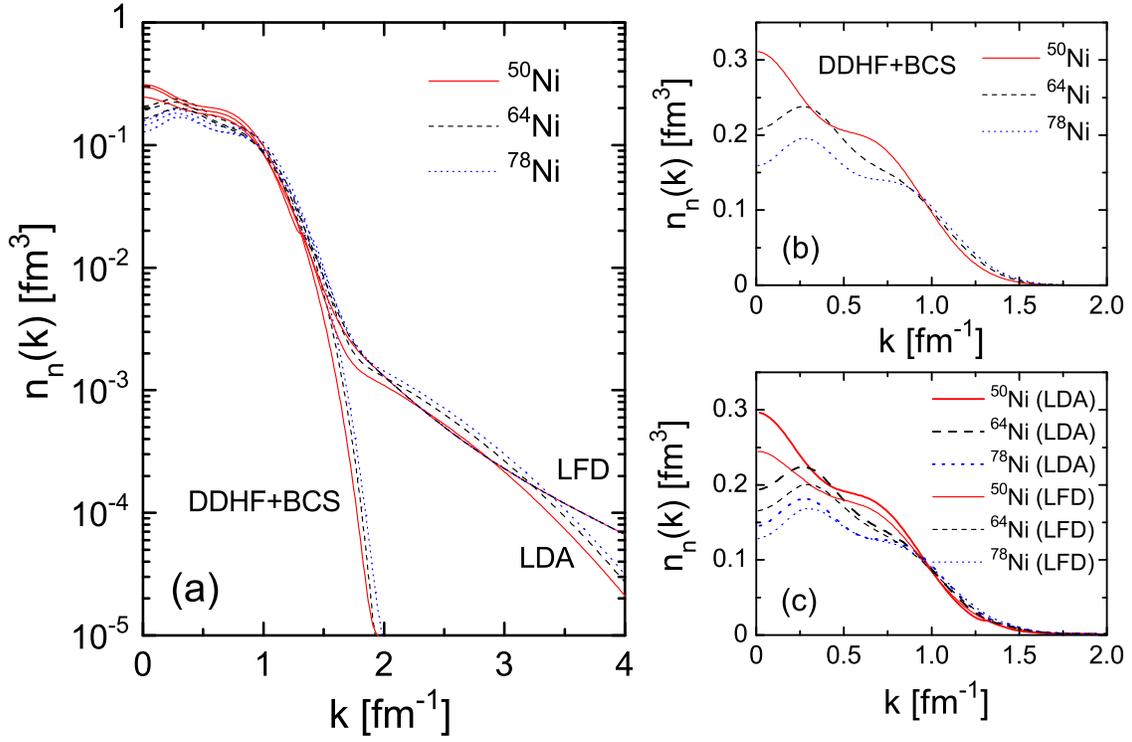}
\caption[]{(a) Neutron momentum distributions obtained within the
DDHF+BCS, LFD, and LDA methods for $^{50}$Ni (solid line),
$^{64}$Ni (dashed line), and $^{78}$Ni (dotted line) isotopes. The
normalization is: $\int n_{n}(k)d{\vec k}=1$. The DDHF+BCS
results, as well as the LFD and LDA results are separately shown
in a linear scale in (b) and (c), respectively. \label{fig4}}
\end{figure*}

\begin{figure*}
\centering
\includegraphics[width=150mm]{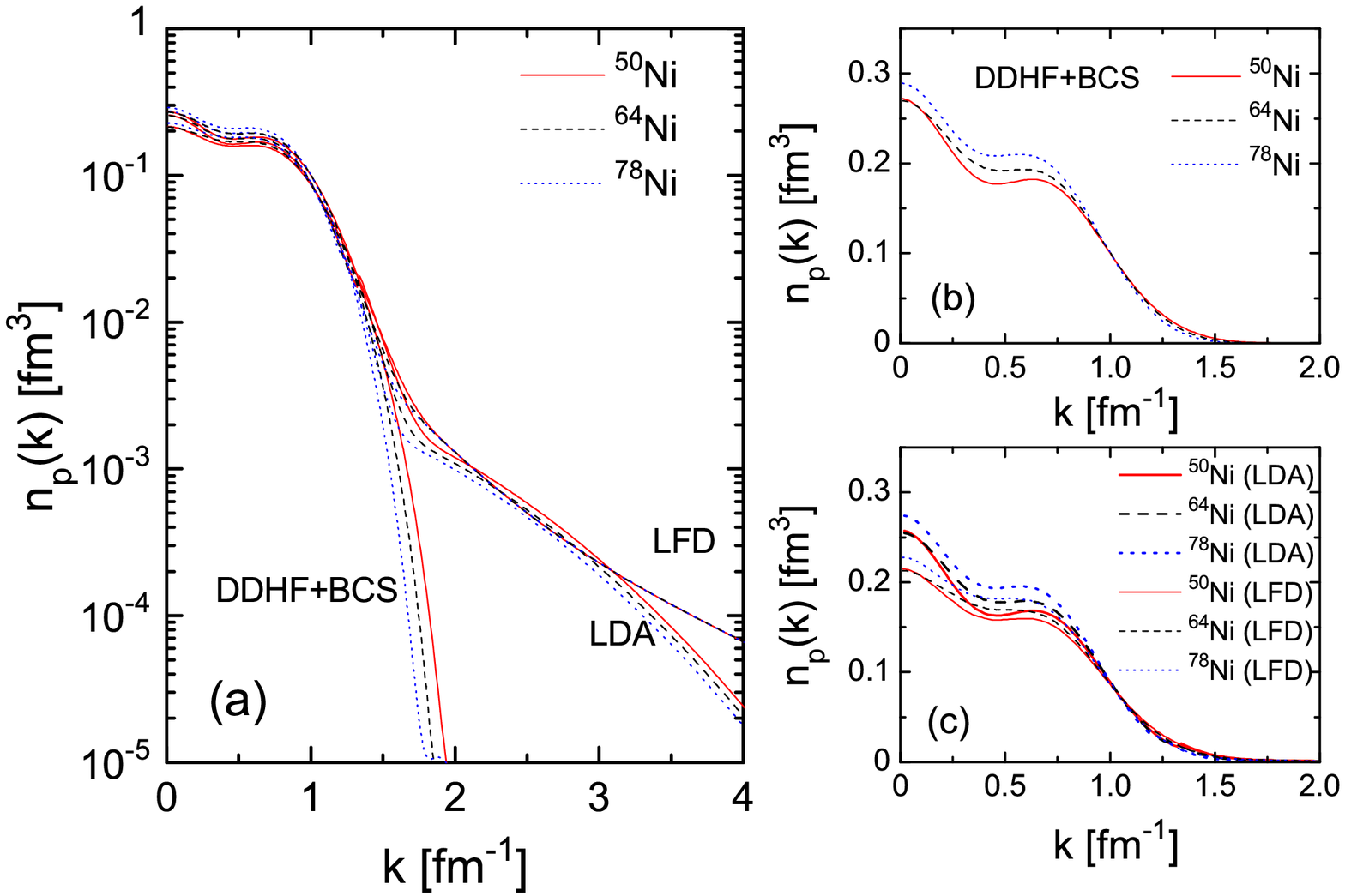}
\caption[]{The same as in Fig.~\ref{fig4}, but for the proton
momentum distributions. \label{fig5}}
\end{figure*}

\begin{figure*}
\centering
\includegraphics[width=150mm]{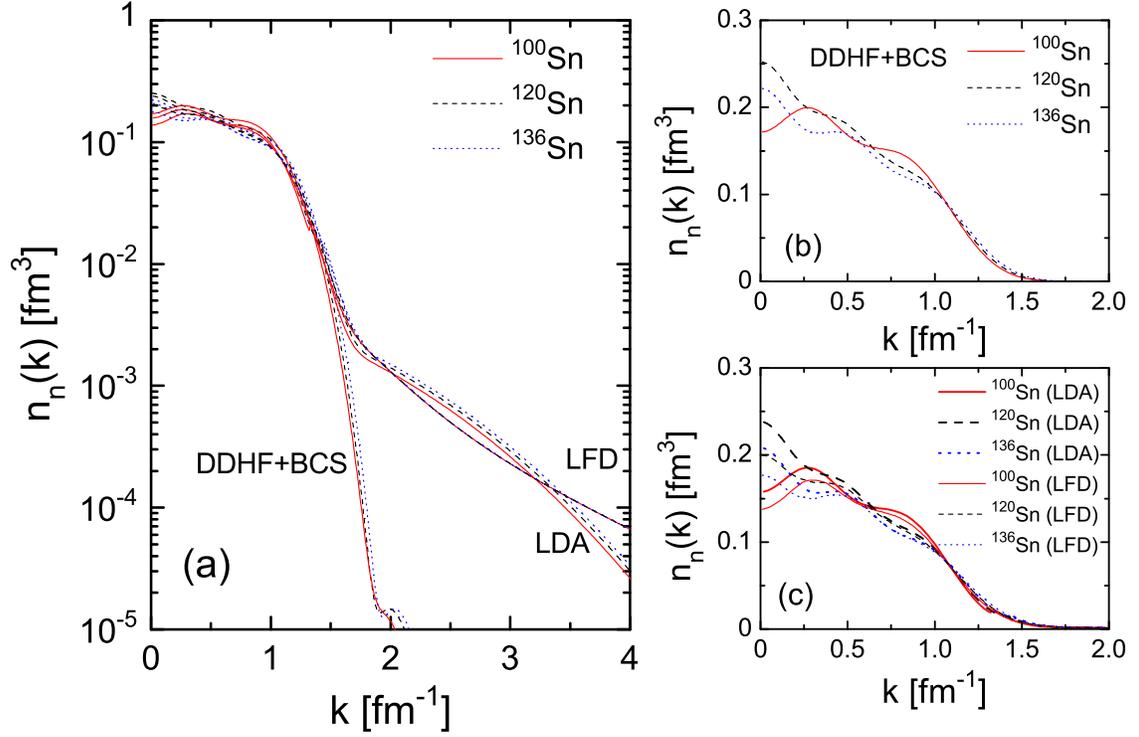}
\caption[]{The same as in Fig.~\ref{fig4}, but for the $^{100}$Sn
(solid line), $^{120}$Sn (dashed line), and $^{136}$Sn (dotted
line) isotopes. \label{fig6}}
\end{figure*}

\begin{figure*}
\centering
\includegraphics[width=150mm]{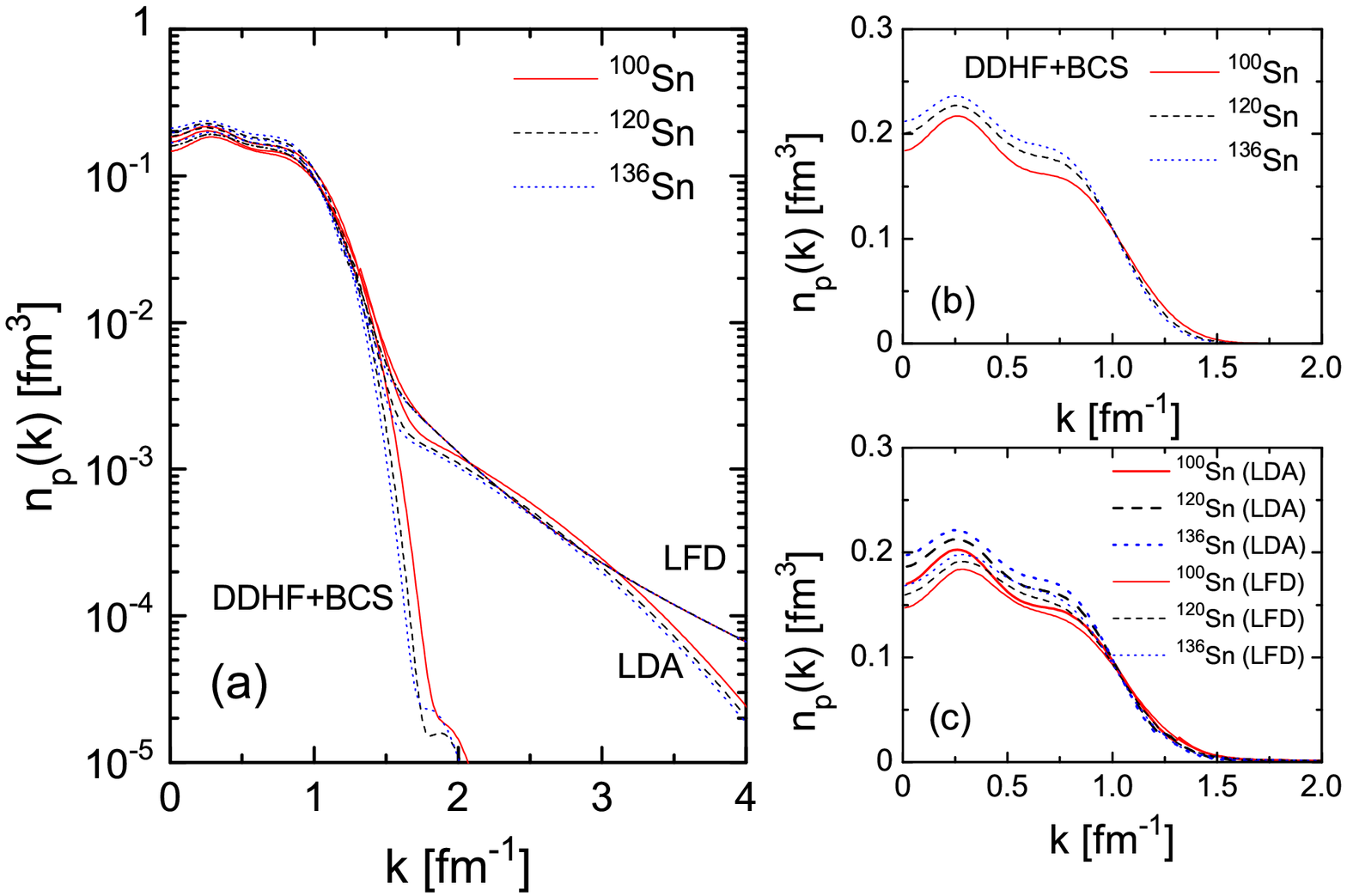}
\caption[]{The same as in Fig.~\ref{fig6}, but for the proton
momentum distributions. \label{fig7}}
\end{figure*}

\begin{figure*}
\centering
\includegraphics[width=170mm]{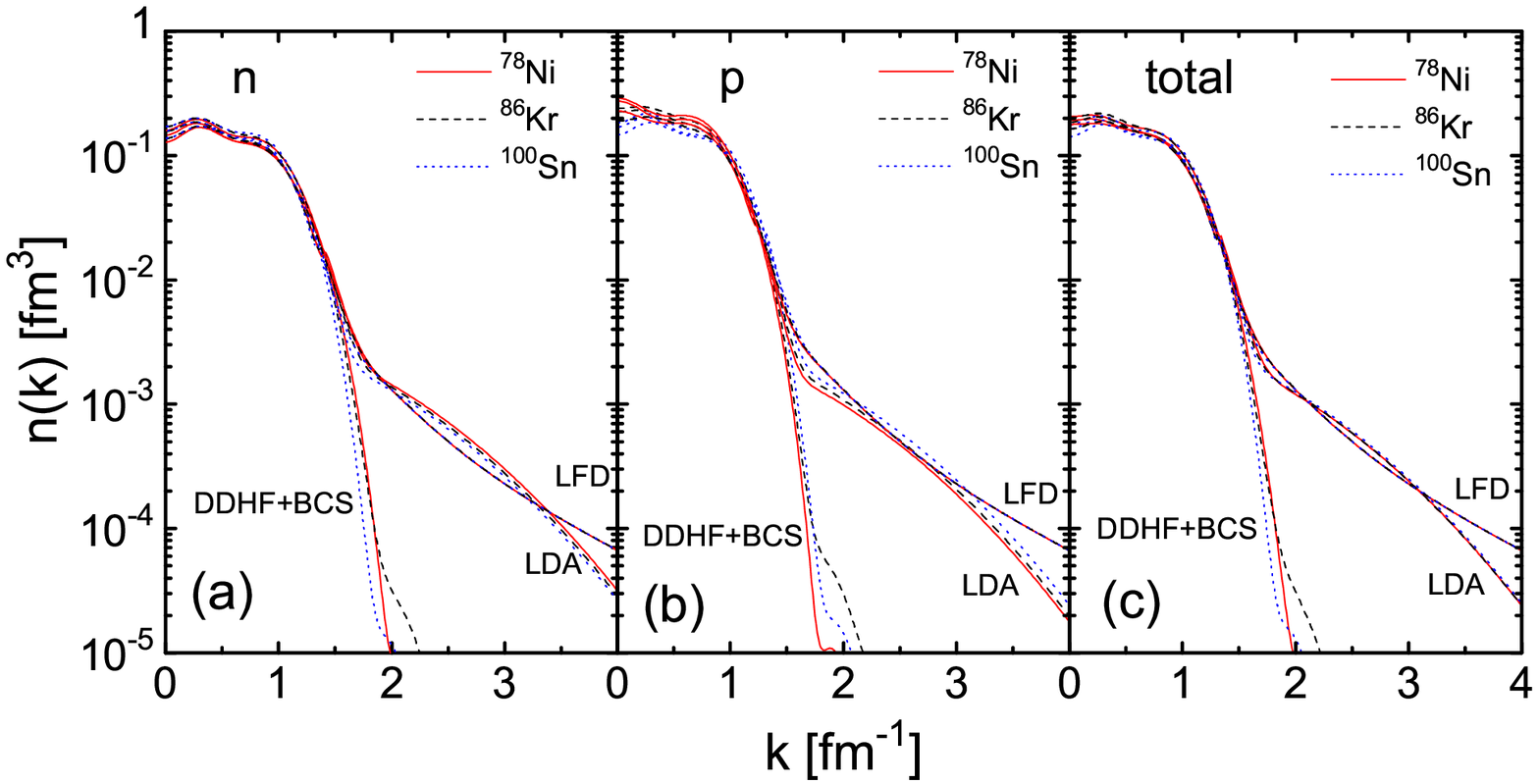}
\caption[]{Neutron (a), proton (b), and total (c) momentum
distributions obtained within the DDHF+BCS, LFD, and LDA methods
for $^{78}$Ni (solid line), $^{86}$Kr (dashed line), and
$^{100}$Sn (dotted line) isotones. The normalization is: $\int
n_{n(p)}(k)d{\vec k}=\int n(k) d{\vec k}=1$. \label{fig8}}
\end{figure*}

The isotonic sensitivity of the neutron, proton and total momentum
distributions of $^{78}$Ni, $^{86}$Kr, and $^{100}$Sn nuclei
($N$=50) is shown in Fig.~\ref{fig8}. It can be seen from the
figure a small difference between the different curves when using
a given approach (DDHF+BCS or LDA). The neutron momentum
distributions $n_{n}(k)$ do not differ so much in comparison to
the proton momentum distributions $n_{p}(k)$. The relative
contributions of $n_{n}(k)$ and $n_{p}(k)$ to the total momentum
distribution $n(k)$ lead to almost equal high-momentum tails of
$n(k)$ for these isotones. This is in accordance with the
well-known results showing that the tails of $n(k)/A$ are almost
equal for all nuclei (for a review, see e.g. \cite{Antbooks}).

\begin{figure*}
\centering
\includegraphics[width=160mm]{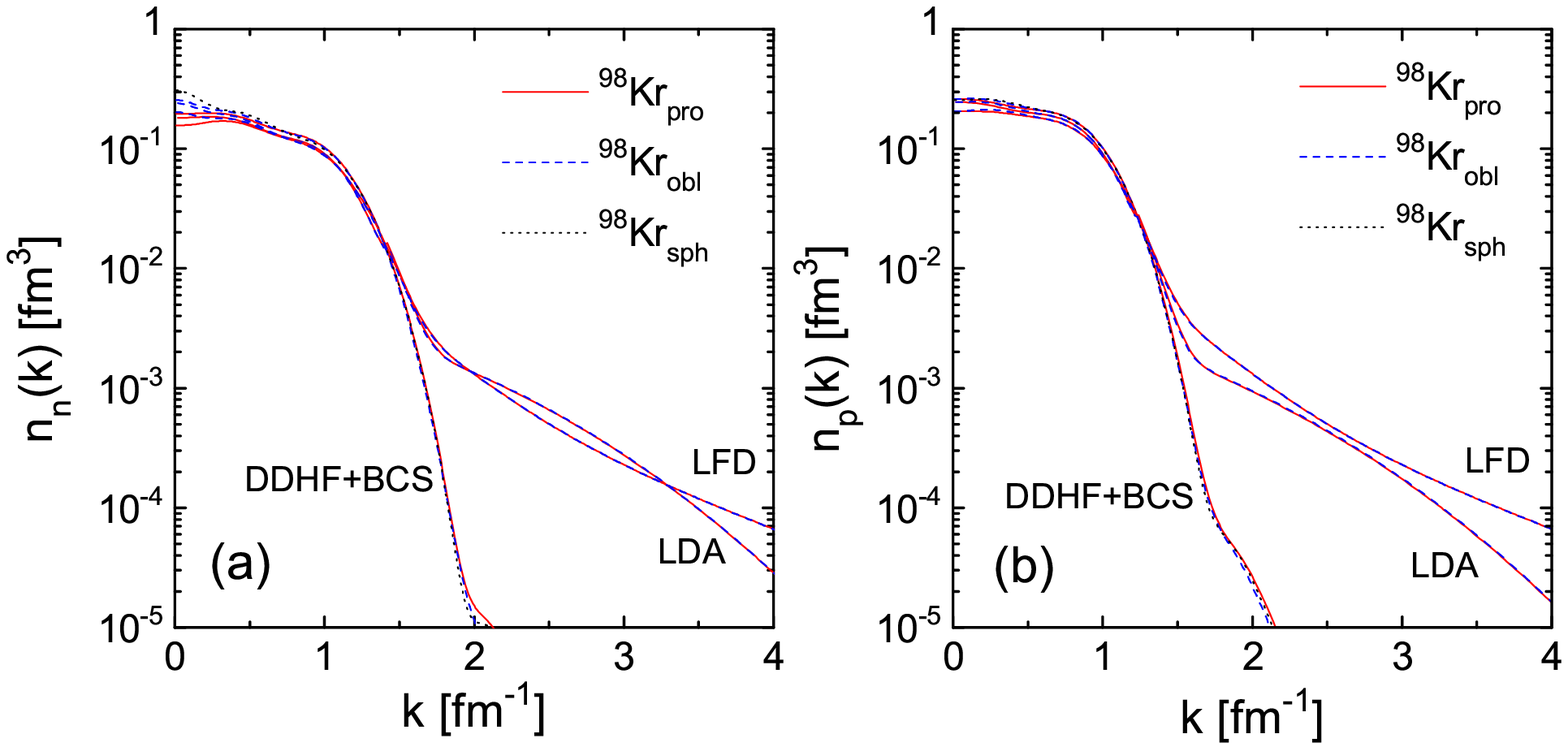}
\caption[]{Neutron (a) and proton (b) momentum distributions
obtained within the DDHF+BCS, LFD, and LDA  methods corresponding
to oblate (dashed line) and prolate (solid line) shape of
$^{98}$Kr. For comparison, the spherical case (dotted line) within
the DDHF+BCS method is also given. The normalization is: $\int
n_{n(p)}(k)d{\vec k}=1$. \label{fig9}}
\end{figure*}

The role of the deformation of the mean field on the NMD is
studied on the example of $^{98}$Kr isotope. For this purpose, we
show in Fig.~\ref{fig9} the intrinsic momentum distribution for
neutrons and protons and for oblate and prolate shapes of
$^{98}$Kr. The results of the three methods considered in our work
are given and compared together with the result for the spherical
case obtained within the HF method. The differences between the
momentum distributions calculated within a given theoretical
method are negligible (especially at high $k$) and practically can
not be distinguished. Thus, we find almost no dependence of the
$n_{n}(k)$ and $n_{p}(k)$ on the character of deformation.

\begin{figure*}
\centering
\includegraphics[width=160mm]{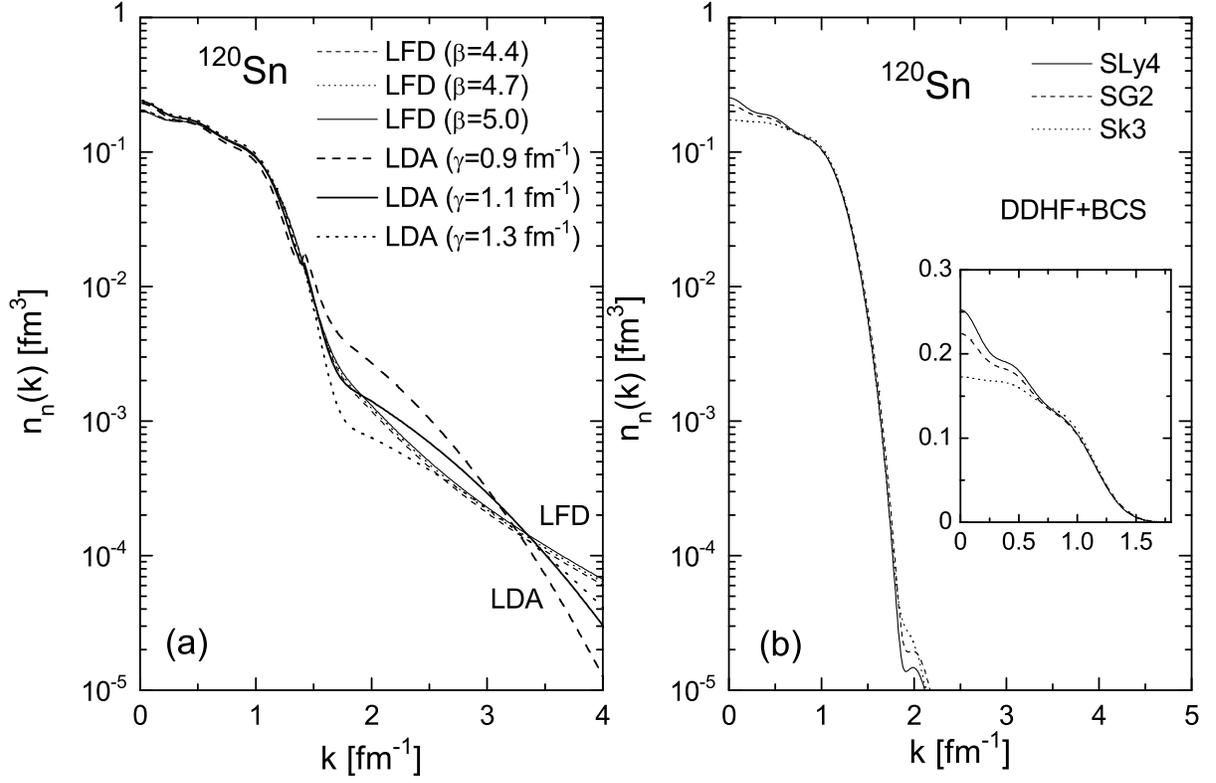}
\caption[]{(a) Neutron momentum distributions obtained within the
LFD (for different values of the parameter $\beta$) and LDA (for
different values of the parameter $\gamma$) methods for
$^{120}$Sn; (b) Neutron momentum distributions obtained within the
DDHF+BCS method for $^{120}$Sn and for different Skyrme forces.
The normalization is: $\int n_{n}(k)d{\vec k}=1$. \label{fig10}}
\end{figure*}

The effects of the correlations included in the LFD and LDA
methods on the neutron momentum distribution of $^{120}$Sn are
presented in Fig.~\ref{fig10}a. The parameters that govern the
correlations in the two approaches are $\beta$ and $\gamma$,
respectively. The results of the calculations for three values of
each of them are shown by thick lines in the case of LDA and by
thin lines in the case of LFD method. The value $\beta$=5 is the
upper limit of this parameter \cite{Ciofi96} and, therefore, we
give in Fig.~\ref{fig10}a also the results for $n_{n}(k)$ for two
smaller values $\beta$=4.4 and $\beta$=4.7. It turns out that
$n_{n}(k)$ does not change significantly for different values of
$\beta$, thus showing the strong presence of correlations at short
distances within the LFD method not only for the stable, but also
for the exotic nuclei. However, a larger sensitivity of $n_{n}(k)$
on the parameter $\gamma$ in the LDA approach appears ,
particularly in the interval $1.5<k<3$ fm$^{-1}$. The value
$\gamma$=1.1 fm$^{-1}$ (from Ref.~\cite{Strin90}) and two more
values $\gamma$=0.9 fm$^{-1}$ and $\gamma$=1.3 fm$^{-1}$ are used
in the calculations. As expected, the obtained neutron momentum
distributions contain smaller correlation effects at larger values
of $\gamma$. This is in accordance with the behavior of the
correlation function $f(r)$ (see Eq.~(\ref{eq:corrfunc})). At
$k>3$ fm$^{-1}$ the LDA results start to deviate in an opposite
way, but in this very high-momentum region no definite conclusion
about the role of correlations can be drawn.

In our work we study also the sensitivity of our results for NMD
to different effective NN forces. In Fig.~\ref{fig10}b we show the
neutron momentum distributions of $^{120}$Sn obtained by using
SLy4 Skyrme force together with the results obtained from other
parametrizations, namely, Sk3 \cite{sk3} and SG2 \cite{sg2}. For
stable spherical nuclei it is known that all the Skyrme-type
effective interactions give similar results for the total momentum
distribution \cite{Casas87}. We would like to resolve possible
ambiguities concerning exotic nuclei by comparing the momentum
distributions for neutrons on the example of the stable $^{120}$Sn
isotope. In fact, some sensitivity to different Skyrme forces can
be observed in the inset of Fig.~\ref{fig10}b for momenta $k<0.5$
fm$^{-1}$. This is due to shell effects that are found also in the
density profiles of the nucleus. Generally, as it has been pointed
out in Ref.~\cite{Sarriguren2007}, the predictions for the charge
and matter rms radii do not differ too much when different Skyrme
forces are used. In analogy, different Skyrme interactions also
produce similar results for the NMD.

\begin{figure*}
\centering
\includegraphics[width=100mm]{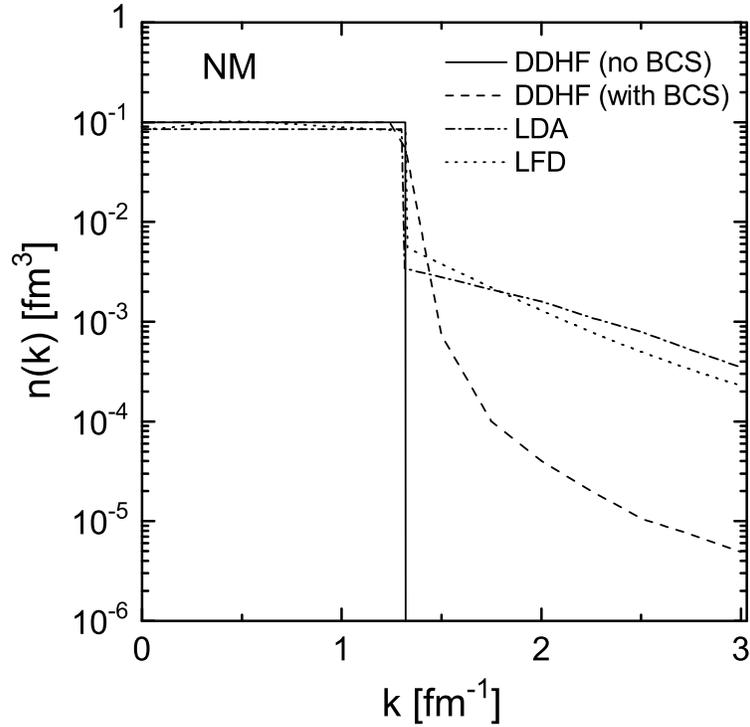}
\caption[]{Comparison of DDHF results for the momentum
distributions of nuclear matter with (dashed line) and without
(solid line) pairing correlations with the results from LFD
(dotted line) and LDA (dash-dotted line) methods. \label{fig11}}
\end{figure*}

Finally, in Fig.~\ref{fig11} a comparison of the results for the
total momentum distribution $n(k)$ of nuclear matter calculated
within the MFA and correlation methods used in our work is shown.
The HF momentum distribution is strongly affected by pairing
correlations which build up a long tail at high momentum
($k>k_{F}$). Comparing this result with the result illustrated in
Fig.~\ref{fig2} for $^{84}$Kr, the different role played by
pairing correlations on the DDHF momentum distribution of NM and
of finite nuclei becomes clear. Moreover, it is interesting to
explore the case when one includes other type of correlations in
NM. Stringari {\it et al.} \cite{Strin90} have already shown in
their model based on the LDA the prediction for $n(k)$ in the case
of nuclear matter. An enhancement of the high-momentum components
of $n(k)$ can be seen from Fig.~\ref{fig11} when both LDA and LFD
methods are used. Hence, in nuclear matter the effects of
short-range and tensor correlations are much stronger than the
BCS-correlations taken into account in the DDHF+BCS calculations.

\section{Conclusions}
\label{summary}

In this work we investigated the properties of the momentum
distributions of medium and heavy exotic nuclei, especially of Ni,
Kr, and Sn even-even isotopes. The theoretical study was performed
on the base of the mean-field method, as well as of two
correlation methods taking into account the NN correlations at
short distances. The neutron, proton, and total momentum
distributions of these nuclei were calculated within: i) a
deformed DDHF+BCS method with Skyrme-type effective interactions
\cite{Sar99,Guerra91}; ii) a theoretical approach
\cite{AGI+02,Antonov2005,Ant2006b} based on the light-front
dynamics method \cite{Carbonell95}; iii) a theoretical model based
on the local density approximation \cite{Strin90}. In the DDHF+BCS
calculations we consider the monopole component of $n(\vec k)$
($n(\vec k)\cong n_{0}(\vec k)$) since this is the only important
component in the expansion of the HF ground-state momentum
distribution [Eq.~(\ref{nmult})]. On the other hand, the two
correlation approaches allow one to include both the mean-field
and short-range effects for the description of the nucleon
momentum distribution.

The study of the isotopic sensitivity of various kinds of momentum
distributions shows different trends. For a given isotopic chain,
we find that in the high-momentum region ($k>1.5$ fm$^{-1}$) the
high-momentum tails of the neutron momentum distributions
$n_{n}(k)$ increase with the increase of the number of neutrons
$N$, while the proton momentum distributions $n_{p}(k)$ exhibit an
opposite effect. In the same region the LFD method does not show
this isotopic sensitivity, in contrast to the DDHF+BCS and LDA
methods. At low momenta $n_{n}(k)$ decreases while, on the
contrary, $n_{p}(k)$ increases with the increase of $N$.
Additionally to the isotopic sensitivity we studied how the
momentum distributions of some isotones are modified keeping the
neutron number constant. We find that the total momentum
distributions of $^{78}$Ni, $^{86}$Kr, and $^{100}$Sn nuclei
($N$=50) reveal the same high-momentum tails in all methods used.

The role of the deformation on the momentum distributions is
discussed in the present work on the example of $^{98}$Kr isotope.
As it has been found in Refs.~\cite{Caballero90,Guerra91}, the
isotropy of the total momentum distribution is a property of the
nucleus at equilibrium. Our results for the neutron and proton
momentum distributions of this nucleus show small changes in the
overall behavior for the oblate and prolate shapes. Although the
neutron and proton densities change with deformation
\cite{Sarriguren2007}, the momentum distributions demonstrate a
very weak dependence on the character of deformation. This is
valid for all three theoretical approaches explored in our work.

The pairing correlations are shown to influence the high-momentum
behavior of the neutron $n_{n}(k)$, proton $n_{p}(k)$, and total
$n(k)$ momentum distributions in the case of $^{84}$Kr, but the
differences between the results with or without BCS-correlations
included in the calculations are very small. The latter are even
smaller considering nuclei from the Ni and Sn isotopic chains that
have a spherical shape. The effect of pairing correlations on the
HF momentum distribution is much stronger in the case of nuclear
matter producing a tail for momenta $k>k_{F}$.

The effects of the dynamical correlations on the momentum
distributions have been calculated using approaches which account
for correlations at short distances. In the correlation approach
based on LFD, the incorporation of the LFD result for the nucleon
momentum distribution in the deuteron makes it possible to study
NMD's also for exotic nuclei, especially at high momenta ($k\geq
2$ fm$^{-1}$). It is known \cite{FCW00,Ciofi96} that the latter
are rescaled versions of $n(k)$ in the deuteron. This is the main
reason for the fact that the LFD calculations do not show isotopic
sensitivity on the obtained NMD's. Concerning the NMD's calculated
in the LDA approach in which a Jastrow-type correlation function
is adopted, some difference between $n(k)$ for protons and
neutrons is observed due to $Z(N)$-dependence of the local Fermi
momentum $k_{F}$. In general, a strong enhancement of the momentum
distributions at large $k$ is observed in comparison to the MFA
result and, at the same time, both LFD and LDA results are similar
in the high-momentum region. In our work we pay a particular
attention to the dependence of the results for the momentum
distributions on the correlation parameters $\beta$ and $\gamma$.
In the LFD calculations we used the value of the parameter
$\beta=5.0$. It turned out that this value leads to a good
agreement with the empirical data for $n(k)$ extracted from the
$y$-scaling analyses of inclusive electron scattering off nuclei
\cite{CW99,CW97}. The LDA calculations have been carried out for
$\gamma$=1.1 fm$^{-1}$, the same value being adopted in nuclear
matter and also providing a good choice for finite nuclei. In
addition, we note that our prediction for $n(k)$ in NM obtained
within the LFD method is in agreement with the LDA result
\cite{Strin90}. In our opinion, however, the question for the
specific values of the parameters $\beta$ and $\gamma$ which
determine the strength of the correlations is still open.

We would like to emphasize that, in our work, a possible practical
way to make predictions for the momentum distributions of exotic
nuclei far from the stability line is proposed that provides a
systematic description of $n(k)$ in medium-weight and heavy
nuclei. The comparison of the predicted nucleon momentum
distributions with the results of possible experiments using a
colliding electron-exotic nuclei storage rings would show the
effect of the neutron excess in these nuclei and will be also a
test of various theoretical models of the structure of exotic
nuclei.

\begin{acknowledgments}
This work was partly supported by the Bulgarian National Science
Fund under Contract No.~02-285, by Ministerio de Ciencia e
Innovaci´on (Spain) under Contract Nos. FIS2008-01301 and
FPA2007-62616, and by the Agreement between CSIC (Spain) and the
Bulgarian Academy of Sciences (2007BG0011). One of the authors
(M.V.I.) acknowledges the support from the European Operational
Program HRD through Contract BGO051PO001/07/3.3-02/53 with the
Bulgarian Ministry of Education, Youth and Science.

\end{acknowledgments}

\end{document}